\documentclass[10pt]{article}

\usepackage{authblk}
\usepackage{natbib}

\usepackage{graphicx}
\usepackage{bm}

\usepackage{amsmath}

\newcommand {\col}[1]{\bm{#1}}
\newcommand {\mat}[1]{\bm{#1}}

\newcommand {\HT}{\dagger}

\newcommand {\s}{\col{\sigma}}
\newcommand {\bl}{\col{u}}

\newcommand {\E}{E}
\newcommand {\C}{C}
\newcommand {\U}{U}
\newcommand {\W}{\Lambda}

\newcommand {\K}{K}

\newcommand {\I}{b}
\newcommand {\D}{d}
\newcommand {\R}{r}

\newcommand {\F}[1]{{\cal{F}}\left\{#1\right\}}
\newcommand {\Fline}[1]{{\cal{F}}\{#1\}}

\newcommand{\appropto}{\mathrel{\vcenter{
  \offinterlineskip\halign{\hfil$##$\cr
  \propto\cr\noalign{\kern2pt}\sim\cr\noalign{\kern-2pt}}}}}

\title{Direction-Dependent Polarised Primary Beams in Wide-Field Synthesis Imaging}

\date{}

\author[1,2]{D. A. Mitchell}
\author[3]{R. B. Wayth}
\author[4]{G. Bernardi}
\author[4]{L. J. Greenhill}
\author[3]{S. M. Ord}

\affil[1]{School of Physics, The University of Melbourne, Parkville, VIC 3010, Australia}
\affil[2]{ARC Centre of Excellence for All-sky Astrophysics (CAASTRO)}
\affil[3]{International Centre for Radio Astronomy Research, Curtin University. GPO Box U1987, Perth WA, 6845. Australia}
\affil[4]{Harvard-Smithsonian Center for Astrophysics, 60 Garden St, Cambridge, MA 02138, USA}

\begin{document}

\maketitle

{\abstract The process of wide-field synthesis imaging is explored, with the aim of understanding the implications of
variable, polarised primary beams for forthcoming Epoch of Reionisation experiments. These experiments seek to detect
weak signatures from redshifted 21cm emission in deep residual datasets, after suppression and subtraction of foreground
emission. Many subtraction algorithms benefit from low side-lobes and polarisation leakage at the outset, and both of
these are intimately linked to how the polarised primary beams are handled. Building on previous contributions from a
number of authors, in which direction-dependent corrections are incorporated into visibility gridding kernels, we
consider the special characteristics of arrays of fixed dipole antennas operating around 100-200 MHz, looking towards
instruments such as the Square Kilometre Array (SKA) and the Hydrogen Epoch of Reionization Arrays (HERA). We show that
integrating snapshots in the image domain can help to produce compact gridding kernels, and also reduce the need to make
complicated polarised leakage corrections during gridding. We also investigate an alternative form for the gridding
kernel that can suppress variations in the direction-dependent weighting of gridded visibilities by 10s of dB, while
maintaining compact support.}


\section{Introduction}

Synthesis imaging in radio astronomy is an indirect technique for measuring the radio sky brightness distribution. An
array of radio receivers is used to measure the interference pattern of celestial radio waves, and the pattern is used
to determine a sky brightness model. See, for example, \cite{Thompson-etal.2001}. However, a number of simplifying
assumptions that are commonly used in radio astronomy (such as small fields of view and identical antennas) are not
valid for many new radio arrays. This is particularly true for new low-frequency arrays of fixed dipoles\,--\,for
example, MWA \citep{Lonsdale-etal.2009, Tingay-etal.2012}, LOFAR \citep{deVos-etal.2009}, PAPER
\citep{Parsons-etal.2010}, and LWA \citep{Ellingson-etal.2009, Taylor-etal.2012}\,--\,where advances in low-cost,
high-performance electronics allow the deployment of many small phased-array antennas. Such antennas have a large field
of view and can have complicated total power and polarisation responses on the sky. Most of these arrays will also
search for the faint, diffuse signal coming from the Epoch of Reionisation \citep[EoR, e.g.,][]{Furlanetto-etal.2006,
MoralesWhiyte2010}, for which foreground subtraction is a critical step \citep[e.g.,][]{Morales-etal.2006}. Recent
observations have already reached the confusion level limit, well above the expected level required to detect the EoR
signal \citep{Bernardi-etal.2009, Bernardi-etal.2010, Williams-etal.2012, Bernardi-etal.2012}. Very high fidelity
imaging is required to accurately model the foreground sky, and this requirement motivates the present work. In the
remainder of this section the problem will be laid out, and in subsequent sections the implications will be explored in
more detail.

Over an infinitesimal range of observing time and frequency, an incident plane wave can be described by a 4-element
column vector containing the self and cross coherence terms of the two electric field polarisations
\citep{Hamaker-etal.1996}. The superposition of many such waves can be described by the spatial coherence of the
electric field, and with enough samples this interference pattern can be inverted to reconstruct the incident power from
the sky. However, when imaging across a wide field of view, angular response variations from sample to sample often
limit the image quality that can be achieved by standard processing packages. There is significant effort underway to
deal with the wide-field problem both in hardware, for instance the addition of a third axis of rotation to ASKAP
antennas,\footnote{See the ASKAP Antenna Public Specification available at
www.atnf.csiro.au/projects/askap/antennas.html.} and in software, by applying direction-dependent corrections to each
sample via convolution in the Fourier domain
\citep[e.g.,][]{Myers-etal.2003,Bhatnagar-etal.2008,MoralesMatejek2009,Smirnov2011b}.

Consider a radio interferometer that measures a complex, 4-element spatial-coherence column vector for each pair of
spatially separated receivers (i.e., a vector containing the four polarised visibility measurements). The angular
response of each receiver pair, or baseline, to each incident plane wave will in general be unique. It will be a
function of receiver separation, the direction-dependent polarised primary beams of the two receivers, including the
direction-dependent transform from sky polarisation coordinates to the polarised inputs of each receiver, and
direction-dependent propagation effects. It is assumed that the signal paths remain linear, allowing the response of a
given baseline to a given direction to be described by a $4\times4$ gain matrix multiplied by a geometric phase term.

For a given baseline, rather than thinking of a separate response matrix for each direction, instead consider a single
$4\times4$ matrix with elements that are two-dimensional angular response functions. Multiplying this matrix with the
angular distribution of radio emission, described as a $4\times1$ vector of two-dimensional brightness functions,
results in a $4\times1$ vector containing the angular distribution of power seen by the baseline. The measured
spatial-coherence vector can be modelled as the integral of these four functions over the field of view, or
alternatively in the Fourier domain using convolutions. The Fourier domain representation is particularly useful for the
reverse direction, going from visibilities to the sky, as it allows for a unique direction-dependent multiplication for
each baseline.

In the following section the situation will be developed mathematically. In general, lower-case bold and italic symbols
will refer to column vectors, while upper-case bold and italic symbols will refer to matrices. Variables that are
functions of position in either the Fourier plane or image plane will be explicitly labeled. 


\section{Estimating Sky Brightness}
\label{Estimating Sky Brightness}

Synthesis imaging involves the inversion of three-dimensional integrals, which are usually reduced to one or more
two-dimensional segments or approximations, such as those described in \cite{Cornwell-etal.2003}. These approaches can
be computationally expensive and are complicated by variable polarisation coordinate systems, both between and within
the two-dimensional segments. As an example, consider making images across the wide field-of-view of an array of dipole
antennas that are fixed to the ground; even with ideal dipoles, polarised emission from radio sources at different
positions on the curved sky will undergo different coordinate transformations, and these transformations will change
with sidereal motion.

Before considering a more complicated brightness distribution, suppose that the sky is comprised of a single point
source in direction $\s^\prime$, with a brightness described by polarised coherency vector, $\col{\I}(\s^\prime)$, each
element being a delta function centred at $\s^\prime$ (with units of W\,m$^{-2}$\,sr$^{-1}$\,Hz$^{-1}$). The $k^{th}$
spatial-coherence sample reduces to

\begin{equation}
\label{DD measurement}
\col{\R}_{k}
 =  \mat{\E}_{k}^{}(\s^\prime)\, \col{\I}(\s^\prime)\, e^{-i2\pi\bl_k.\s^\prime} + \col{n}_{k}^{},
\end{equation}

where $\mat{\E}_{k}^{}(\s^\prime)$ is a $4\times4$ matrix that describes the angular response of the sample to the sky,
$\bl_k$ is the position vector of the sample's baseline measured in wavelengths, such that the exponent represents a
geometric phase shift imposed by the baseline, and $\col{n}_{k}$ is a $4\times1$ column vector of independent
measurement noise. From $N$ such measurements, and assuming Gaussian-distributed measurement noise, a maximum-likelihood
estimate of the incident coherency vector can be made using the normal equations \citep[e.g.,][]{Press1986}:

\begin{equation}
\label{DD single source MLE}
\hat{\col{\I}}(\s^\prime)
  = \left( \sum_{k=1}^N \mat{\E}_k(\s^\prime)^\HT \, \mat{\W}_{k}^{} \, \mat{\E}_k^{}(\s^\prime) \right)^{-1}
    \left( \sum_{k=1}^N \mat{\E}_k(\s^\prime)^\HT \, \mat{\W}_{k}^{} \,
                  \col{\R}_k^{} \, e^{i2\pi\bl_k.\s^\prime} \right) = \mat{D}(\s^\prime)^{-1}\col{\D}(\s^\prime),
\end{equation}

where the inverse-variance weight matrix, $\mat{\W}_{k}$, is the inverse of the expectation of
$\col{n}^{}_{k}\col{n}^\HT_{k}$, $\mat{D}$ and $\col{\D}$ represent the bracketed terms, and superscript $\HT$ denotes a
conjugate transpose. See \cite{Ord-etal.2010} and \cite{Bernardi-etal.2012} for practical examples.

Consider now the more complicated case of simultaneously estimating the brightness in $M$ directions, where the
spatial-coherence vector for baseline $k$ is

\begin{equation}
\label{FT measurement}
\col{\R}_{k} = \sum_{m=1}^M \mat{\E}_{k}^{}(\s_m)\, \col{\I}(\s_m)\, e^{-i2\pi\bl_k.\s_m} + \col{n}_{k}^{}.
\end{equation}

Equation (\ref{DD single source MLE}) can be expanded for the $4M\times1$ solution vector
$[\hat{\col{\I}}(\s_1),\hat{\col{\I}}(\s_2),\ldots,\hat{\col{\I}}(\s_M)]^T$, where the right-hand bracketed term is
replaced by $[\col{\D}(\s_1),\col{\D}(\s_2),\ldots,\col{\D}(\s_M)]^T$, a $4M\times1$ vector of dirty measurements in the
instrument frame, and the left-hand bracketed term is replaced by the inverse of the $4M\times4M$ normal equation
matrix, which contains the covariances of the incident coherency estimates.\footnote{The $4M\times4M$ matrix being
inverted is typically large, ill-conditioned or singular, and limited by pixelisation effects. Iterative deconvolution
methods can in general address all of these issues for radio sources with emission that is sufficiently strong.} The
normal equation matrix will contain the $4\times4$ direction-dependent matrices defined in (\ref{DD single source MLE})
along the diagonal, $\mat{D}(\s_1),\mat{D}(\s_2),\ldots$, with polarised point spread function (PSF) side-lobes from
other directions in off-diagonal elements. If the side-lobes of the array are small\,--\,or, for example, if they are
being decreased in an iterative deconvolution process, as in \cite{Bhatnagar-etal.2008}\,--\,the matrix will be
approximately block diagonal, allowing each direction to be approximately transformed to sky coordinates using (\ref{DD
single source MLE}). The remainder of this discussion will be limited to such local transformations, rather than full
deconvolutions, and we will use the subscript $d$ to indicate that a measurement, $\hat{\col{\I}}_d(\s^\prime)$, is
potentially corrupted by the side-lobes of other sources (i.e., is dirty).

For each sample, $k$, (\ref{FT measurement}) has the form of a Fourier transform. It is common with imaging arrays to
arrange the $M$ directions of interest on a regular two-dimensional grid, as it is often computationally efficient to
grid the samples onto the appropriate grid in Fourier space and use the Fast Fourier Transform (FFT) algorithm to
calculate the $M$ measurements, $\col{\D}(\s)$. When $\mat{\E}_{k}(\s)$ is constant for all of the samples, the Fourier
relationship holds for all $k$ and the angular response of the samples can be dealt with after imaging. In addition,
deconvolution is simplified because the PSF is shift-invariant. This is a common assumption in radio interferometry
\citep{Thompson-etal.2001}. In general, however, the spatial-coherence samples will have different sky response
matrices\,--\,for instance, due to different antenna responses or source motion across the sky. Any differences that
affect the entire field of view in a consistent way (such as a time-dependent rotation of the parallactic angle for
alt-az mounted dishes) can be dealt with using simple transformations prior to gridding, but for wide fields of view or
when high dynamic range measurements are required, samples may require direction-dependent transformations to maintain
the Fourier relationship.

When $\mat{\E}_{k}^{}(\s)$ does need to be accounted for on a sample-by-sample basis, direction-dependent
multiplications can be applied during the FFT gridding step, as discussed in previous work
\citep{Bhatnagar-etal.2008,Rau-etal.2009,MoralesMatejek2009}. Let $\mat{\U}_{k}^{}(\bl)$ be the matrix with elements
that are the 2D Fourier transforms of the elements of $\mat{\E}_{k}^{}(\s)$, and consider polarisation product $j$ of
vector $\col{\D}(\s^\prime)$. Writing the Fourier transform operations as integrals, to distinguish them from the
gridding operations, and keeping in mind that the elements of $\col{\R}_{k}$ are samples, not functions of $\bl$, we
have

\begin{equation}\begin{array}{rcl}
\label{MLE imaging}
\D_j(\s^\prime)
  =  \displaystyle\sum_{k=1}^N \sum_{i=1}^4 \frac{1}{\sigma_{k,i}^2}\, \E_{k,ij}^*(\s^\prime)\,
                                            e^{i2\pi\bl_k.\s^\prime}\, \R_{k,i}
& = & \displaystyle\sum_{k=1}^N \sum_{i=1}^4 \frac{1}{\sigma_{k,i}^2}\,
      \iint\limits_{\;\;-\infty}^{\;\;\;\;\;\infty}\!
             \left\{ \U_{k,ij}^*(\bl-\bl_k)\right\} e^{i2\pi\bl.\s^\prime} \mathrm{d}\bl
      \, \R_{k,i}\\
&&\\
& = & \displaystyle\iint\limits_{\;\;-\infty}^{\;\;\;\;\;\infty}\! \left\{
	     \sum_{k=1}^N \sum_{i=1}^4 \frac{1}{\sigma_{k,i}^2}\, \U_{k,ij}^*(\bl-\bl_k)\, \R_{k,i}
      \right\} e^{i2\pi\bl.\s^\prime} \mathrm{d}\bl.
\end{array}\end{equation}

This shows that, as desired, we can grid using kernels that are based on the FFT of the sample response functions, and
form dirty images that are mathematically equivalent to the pixel-by-pixel direct Fourier transforms shown in (\ref{DD
single source MLE}).

\cite{Bhatnagar-etal.2008} consider applications where the $\mat{\E}_{k}(\s)$ matrices are approximately unitary in
every direction of interest (or, less stringently, where $\mat{\E}_{k}(\s)^\HT$ is approximately a scaled version of
$\mat{\E}_{k}(\s)^{-1}$). As they discuss, in a deconvolution scheme where one subtracts a sky model before gridding,
high quality deconvolution comes from accurately estimating and subtracting the model visibilities. During the
subsequent gridding and imaging step, the operations in (\ref{MLE imaging}) will produce nominally calibrated images in
sky polarisation coordinates, due to the approximate unitarity of each $\mat{\E}_{k}(\s)$. While not perfectly
calibrated, these images will in many cases be good enough to incrementally update the sky model. However, the argument
can be made more generally for any $\mat{\E}_{k}(\s)$, as long as the appropriate $\mat{D}(\s)^{-1}$ is used to
transform the dirty images back to sky coordinates before updating the sky model. In general there is not a one-to-one
mapping between the sky, the visibilities, and the dirty images, and to use all of the available information all four
spatial-coherency polarisations will need to be gridded into each of the four instrument polarisation images, as shown
in (\ref{MLE imaging}).

For practical reasons it is typically assumed that, after standard direction-independent calibration, each
$\mat{\E}_{k}(\s)$ matrix has negligible off-diagonal elements, maintaining a one-to-one mapping between the sky and
instrument coherence vectors. It is also typical to require that elements of $\mat{\U}_{k}^{}(\bl)$ have compact support
(that is, are non-zero only for small $\bl$), which is a reasonable assumption given the limited collecting area of
antennas.\footnote{The 3D nature of some antennas will result in power being distributed across 2D Fourier space, but
for the purposes of this paper we will assume the effect is negligible.} These requirements are important, as the
computational cost of gridding with large convolution kernels can be prohibitive. While much of this comes down to
antenna design and engineering tolerances, small-scale structure in the celestial coordinates, particularly for large
fields-of-view and fields near the celestial poles, will lead to large-scale features and off-diagonal terms in
$\mat{\U}_{k}^{}(\bl)$.

If we instead grid to a coordinate system that is aligned with the receivers, such as one aligned with the ideal
polarisation axes in a linearly polarised receiving system, much of the small-scale structure can be dealt with in the
image plane. Furthermore, gridding in an antenna-based coordinate system should reduce the amount of information in the
off-diagonal terms. Gridding in such a coordinate system is natural for arrays with antennas that track the sky, but
this is not always feasible or desirable. It is also natural for snapshot imaging approaches that transform to the image
plane regularly, as described later in this section.

If the angular response of sample $k$ within an antenna-based frame is denoted $\col{\E}_{Ak}(\s)$, such that
$\col{\E}_{k}(\s)=\col{\E}_{Ak}(\s)\,\mat{\C}(\s)$, and if the feed configuration matrix that transforms between frames,
$\mat{\C}(\s)$, is constant across the samples, then one can form an alternative dirty image

\begin{equation}
\label{MLE local}
\col{\D}_A(\s)
 = \sum_{k=1}^N \mat{\E}_{Ak}(\s)^\HT\, \mat{\W}_{k}^{}\, e^{i2\pi\bl_k.\s}\, \col{\R}_{k}
 = \F{ \sum_{k=1}^N \col{\U}_{Ak}(\bl-\bl_k)^\HT\, \mat{\W}_{k}^{}\, \col{\R}_{k} },
\end{equation}

where $\F{\ldots}$ denotes the element-by-element Fourier transform described above for $\mat{\U}(\bl)$ and
$\mat{\E}(\s)$. Substituting for $\col{\E}_{k}(\s)$ in (\ref{DD single source MLE})
results in a single transformation to sky coordinates, applied in the image plane:

\begin{equation}
\label{DD local MLE}
\hat{\col{\I}}_d(\s)
 = \mat{\C}_{}(\s)^{-1}\,\left( \sum_{k=1}^N \mat{\E}_{Ak}(\s)^\HT\, \mat{\W}_{k}^{} \, \mat{\E}_{Ak}^{}(\s) \right)^{-1}
   \col{\D}_A(\s),
\end{equation}

where the subscript $d$ indicates that the estimate is dirty, in that while each pixel has been transformed to sky
coordinates and de-weighted, the pixel coupling contained in the ${4M}\times{4M}$ normal equation matrix has not been
dealt with. Figure \ref{plot_4x4_tile_beams_paper} gives examples of these functions for MWA-style antennas with various
calibration errors.

When tracking a patch of sky with some types of steerable antennas, matrix $\mat{\C}(\s)$ will not change with time,
except for a possible parallactic rotation that is the same for each $\s$ and therefore straightforward to account for
in the Fourier domain. The situation is quite different for an array of dipole antennas that are fixed to the ground.
The structure in $\mat{\C}(\s)$ will change with time in a direction-dependent fashion, becoming more pronounced as the
field size increases. These changes will need to be incorporated into the $\col{\U}_{Ak}(\bl)$ matrices, potentially
increasing the size of the gridding kernels and the relevance of the off-diagonal terms. The longer one integrates, the
more the matrices $\mat{\C}(\s)$ will change, and in general there will be a trade-off between the time interval over
which one can integrate visibility samples and the compactness of $\col{\U}_{Ak}(\bl)$. This situation is alleviated in
snapshot imaging, where much of the structure can be dealt with from snapshot to snapshot in the image plane, using a
time-dependent matrix, $\mat{\C}_t(\s)$. In this case (\ref{DD single source MLE}) and (\ref{MLE local}) can be combined
to give

\begin{equation}
\label{D and T dep local MLE}
\hat{\col{\I}}_d(\s)
 = \left( \sum_{t=1}^T\sum_{k=1}^N \mat{\E}_{kt}(\s)^\HT \, \mat{\W}_{kt}^{} \, \mat{\E}_{kt}^{}(\s) \right)^{-1}
    \sum_{t=1}^T\mat{\C}_t(\s)^\HT\,
                \F{ \sum_{k=1}^N \col{\U}_{Akt}(\bl-\bl_k)^\HT\, \mat{\W}_{kt}^{}\, \col{\R}_{kt} }.
\end{equation}

Snapshot imaging has been avoided in the past due to computational concerns and the desire to enhance deconvolution by
maintaining a constant PSF during an observation \citep[e.g.,][]{CornwellPerley1992,SynImRAII-Ch19}. However, it should
be clear from equation \ref{DD single source MLE} and onwards that in general the convolution theorem will not hold, and
the PSF will vary with both direction and time, regardless of the wide-field imaging strategy. Snapshot imaging software
has been developed to image data from arrays such as the MWA and the Large Aperture Experiment to Detect the Dark Ages
\citep[LEDA, see][]{GreenhillBernardi2012}, enabling image-based corrections for most of the polarisation structure, for
wide-field w-terms, and for direction-dependent position shifts and Faraday rotation caused by the ionosphere
\citep[see, for example,][]{Mitchell-etal.2008, Ord-etal.2010, Tingay-etal.2012}. These arrays are also being designed
to have dense sampling in $\bl$, giving more control of the PSF and helping to minimise the drawbacks of joint
deconvolution.\footnote{Since the convolution theorem does not in general hold, the term deconvolution is a misnomer.
However, as it is used more generally in radio astronomy we will stick with the jargon in this paper.} Developments are
also being made on image-domain deconvolution approaches, which allow for direction- and time-dependent PSFs
\citep[]{Pindor-etal.2011, Bernardi-etal.2011, Williams-etal.2012, Sullivan-etal.2012}.

Before moving on to discuss these concepts further, a few of them should be reiterated. Any direction-dependent
structure in the angular response matrices that is constant for all samples can be dealt with once, after imaging.
Similarly, components of the matrices that are constant across the image can be dealt with before visibility gridding.
However, direction-dependent structure that differs from sample to sample may require direction-dependent modifications
during gridding, using Fourier transforms of the angular response functions. Gridding is computationally expensive, so
it is desirable for the angular response functions to have compact support in Fourier space, to minimise the gridding
kernel size. Furthermore, while in many situations the polarised angular response matrices can be assumed to be
diagonal, resulting in a one-to-one mapping between the sky, the visibilities, and the dirty images, in general this is
not the case. Any non-negligible off-diagonal terms result in extra gridding, so it is desirable to suppress them where
possible. Dipole arrays are precisely the type with substantial off-diagonal elements. The wide-field arrays coming
on-line this decade will provide important for understanding how susceptible various array designs are to these
problems.

Gridding in an antenna-based coordinate system, rather than a sky-based system, can help to both reduce the off-diagonal
terms of the polarised angular response matrices and keep the gridding kernels compact. However, for arrays of dipoles
that are fixed to the ground, the sky coordinates will rotate in a direction-dependent fashion within an antenna-based
system, and there will be a trade-off between integration time in Fourier space and both kernel size and off-diagonal
amplitudes. Snapshot imaging approaches\,--\,which are being developed for arrays of many small antennas, to deal with
ionospheric distortions, wide-field effects, and ever higher bit rates from correlators\,--\,transform samples to the
image plane regularly, allowing much of the time-dependent variability to be accounted for in the image plane and
keeping the kernels compact.

Some of these concepts are shown graphically in section \ref{Examples}, along with the concepts that are discussed
below.


\section{Alternative Gridding Kernels}
\label{Alternative Gridding Kernels}

In radio astronomy it is common to weight and taper the visibility samples being gridded, to control properties of the
PSF and image noise. Two standard weighting types are natural weighting, where each sample is given the same weight (or
in some schemes weights proportional to the inverse of the sample noise variance), and uniform weighting, where the
local density of samples is divided out (or the local density of weights, if visibilities are inverse-variance
weighted), giving different regions of Fourier space equal weight. Under the common assumptions of traditional synthesis
imaging, natural weighting gives the lowest image noise, while uniform weighting reduces PSF side-lobes, since in effect
it is normalising the Fourier coefficients. Schemes such as robust weighting try to get the best of both worlds, for
instance by uniform weighting and then applying a smooth taper that roughly follows the gross distribution of samples
(i.e., retaining much of the low-noise weighting but removing small-scale weight variations that will ring in image
space).

The direction-dependent gridding described in the previous section is effectively an extension of natural weighting,
taken to the extreme of accounting for gain differences of different samples in different directions. As in natural
weighting, the aim is to minimise thermal noise in the image, given adequate deconvolution. While one can attempt to
apply robust weighting to lower the side-lobes, it is also worth asking whether alternative gridding kernels can be used
as a direction-dependent extension to uniform weighting, as one might expect that normalising the Fourier coefficients
in a direction and sample dependent fashion might help to suppress structure (or structure variation) in the PSF.

Naively, one might hope to use the inverse of the sample response matrix, $\F{\mat{\E}_{k}(\s)^{-1}}$. This has a major
drawback: discontinuities in the polarised response, such as nulls, will become sharp features that will ring out across
Fourier space and result in large, complicated gridding kernels. The noise properties of the gridded visibilities will
also be undesirable. An alternative gridding function that shows some promise for situations where the
$\mat{\E}_{k}(\s)$ matrices can be considered diagonal (so that any primary beam discontinuities are likely to be
approximately aligned, which is not the case for cross-polarisations) is

\begin{equation}
\label{Antidote Kernel}
\K_{k,ii}(\bl) = \F{ \bar{\E}_{ii}^*(\s)\,\bar{\E}_{ii}^{}(\s) \, \left( \E_{k,ii}^{}(\s)+\epsilon \right)^{-1} },
\end{equation}

where subscript $ii$ indicates diagonal element $i$, the overbar indicates an average or ideal model, and $\epsilon$ is
used to limit the depth of nulls in the denominator. While the kernels discussed in section \ref{Estimating Sky
Brightness} result in gridded visibilities that are weighted by the square of the angular response, this type of kernel
aims to replace sample-to-sample variations and weight by the square of the average angular response. This has various
benefits:

\begin{itemize}

  \item Deep nulls in $\bar{\E}_{ii}^*(\s)\,\bar{\E}_{ii}^{}(\s)$ will typically suppress areas where there are nulls in
	$\E_{k,ii}(\s)$, and will therefore also suppress ringing in the kernels.

  \item If discontinuities arising from the inverse in (\ref{Antidote Kernel}) are avoided, the gridding kernel should
	have comparable compact support to that of $\F{\bar{\E}_{ii}^*(\s)}$. This is shown for simulated array elements
	comprising a 4$\times$4 grid of dipoles in figure \ref{plot_4x4_tile_beams_extra}.

  \item The gridded samples retain the average direction-dependent weighting, which will often be more significant than
	sample-to-sample variations.

  \item Except for a possible pixel-by-pixel transformation from an antenna-based reference frame to the sky frame, the
	resulting images are correctly weighted for integration in the image domain, for instance when integrating
	snapshots or performing image-domain mosaicing.

  \item $\bar{\E}_{ii}^*(\s)\,\bar{\E}_{ii}^{}(\s)$ is constant across all of the samples, which should in general
	reduce structure and direction dependence in the PSF. It can also simplify weighting: the natural and uniform
	weighting schemes discussed above are usually based on the assumption that the samples have equal gain, but the
	kernels discussed in section \ref{Estimating Sky Brightness} result in gridded samples that have
	direction-dependent gain, and the normalisation for one direction may not be right for other directions.

  \item Over a short enough bandwidth, one could choose $\bar{\E}_{ii}^*(\s)\,\bar{\E}_{ii}^{}(\s)$ to be independent of
	frequency, in an attempt to remove frequency-dependent structure from the PSF. While this would not result in
	the optimal weighting required to minimise thermal noise after image integration, it may prove beneficial for
	some of the foreground-subtraction procedures being researched for 21cm power spectrum experiments \citep[such
	as those described in][]{Bowman-etal.2009}.

\end{itemize}

The details of how these alternative kernels compare with those of section \ref{Estimating Sky Brightness} depend on
issues such as the type of antenna, the level of primary beam variation, the density of samples, and the type of
weighting applied before the FFT. The authors are investigating whether both kernel types might have a place in an
iterative deconvolution and imaging scheme; for instance, using the kernels from section \ref{Estimating Sky Brightness}
while optimising the sky models, to make the most of signal-to-noise ratios, and using the alternative kernels discussed
above in the final iteration to reduce the side-lobes and position-dependence of the PSF in the residual images used for
EoR observations. This work will be the basis of a follow-up paper.


\section{Examples}
\label{Examples}

\begin{figure}[ht]
\begin{center}
\includegraphics[width=16.5cm]{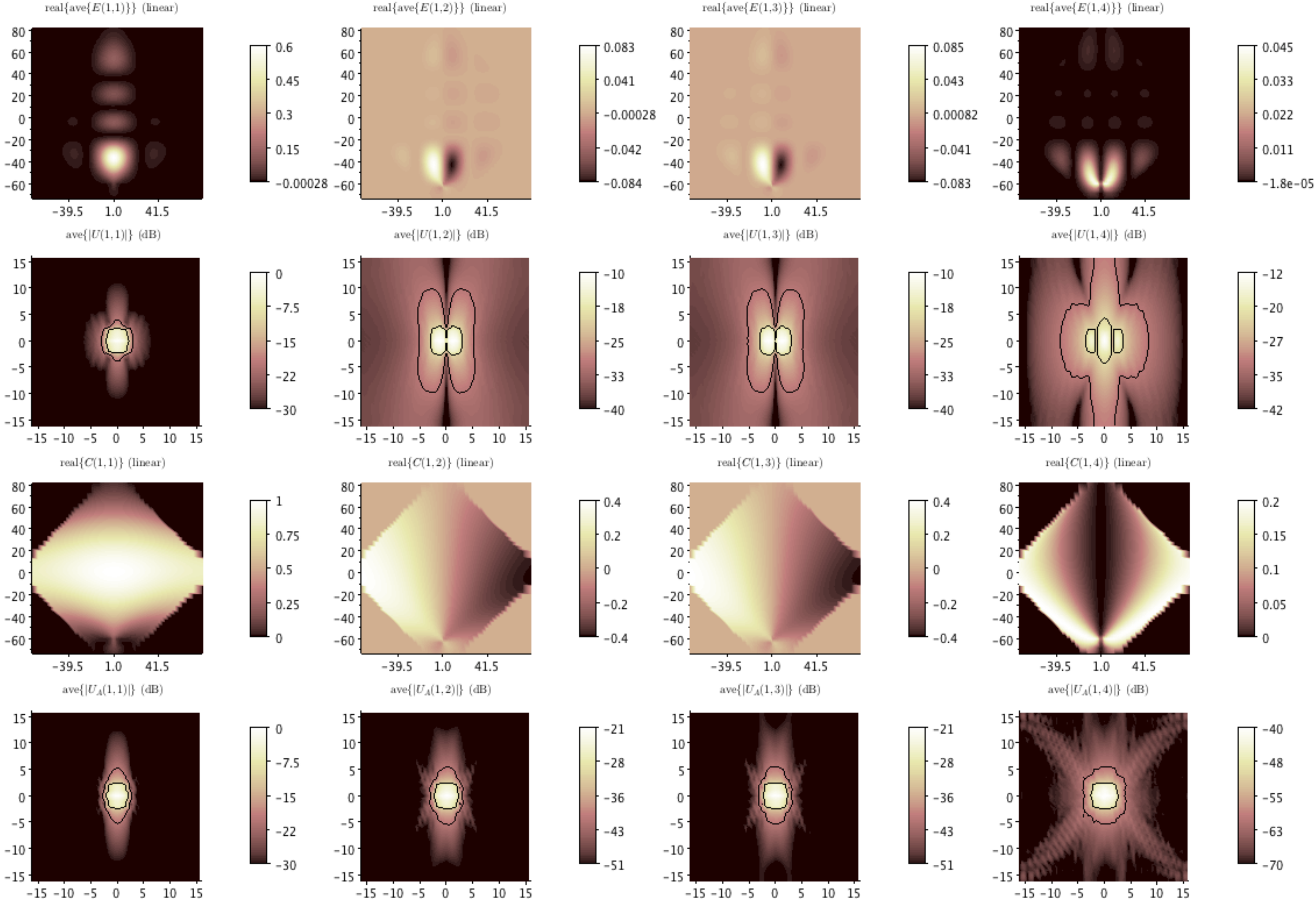}
\end{center}

\caption{The panels show all-sky angular response functions and their FFTs for MWA-style antenna tiles, averaged over 50
	 samples in a single snapshot. They demonstrate how moving to an antenna-based coordinate system reduces the
	 amplitude and extent of the off-diagonal FFT terms, as described in section \ref{Estimating Sky Brightness}.
	 Reading down each column, we have elements of $\mat{\E}_k(\s)$, $\mat{\U}_k(\bl)$, $\mat{\C}(\s)$, and
	 $\mat{\U}_{Ak}(\bl)$. Each column represents one of the 16 matrix elements, these four being those associated
	 with the first element of each spatial-coherency sample. In the image domain, the real components are shown,
	 with linear amplitude scaling and $\s$ in degrees. The structure towards the bottom of the images is due to the
	 south celestial pole. In the Fourier domain, averages are over the absolute values, functions have logarithmic
	 amplitude scaling (running from -30 dB to 0 dB relative to the panel maximum, all scaled to the maximum of
	 element 1,1), and $\bl$ is in wavelengths. Tile primary beams are modelled as the superposition of 16 dipole
	 angular response matrices with various dipole-based errors, as described in the main text, and these
	 imperfections set the final level of leakage that is achieved in an antenna-based coordinate system, shown in
	 the bottom row of panels. The high level of leakage seen for the sky-based coordinate system, shown in the
	 second row, which is as much as 10\% of the amplitude in the 1,1 diagonal element and extends across Fourier
	 space, is due to the structure in $\mat{\C}(\s)$, and highlights the benefit of gridding in an antenna-based
	 coordinate system.}

\label{plot_4x4_tile_beams_paper}
\end{figure}

\begin{figure}[ht]
\begin{center}
\includegraphics[width=16.5cm]{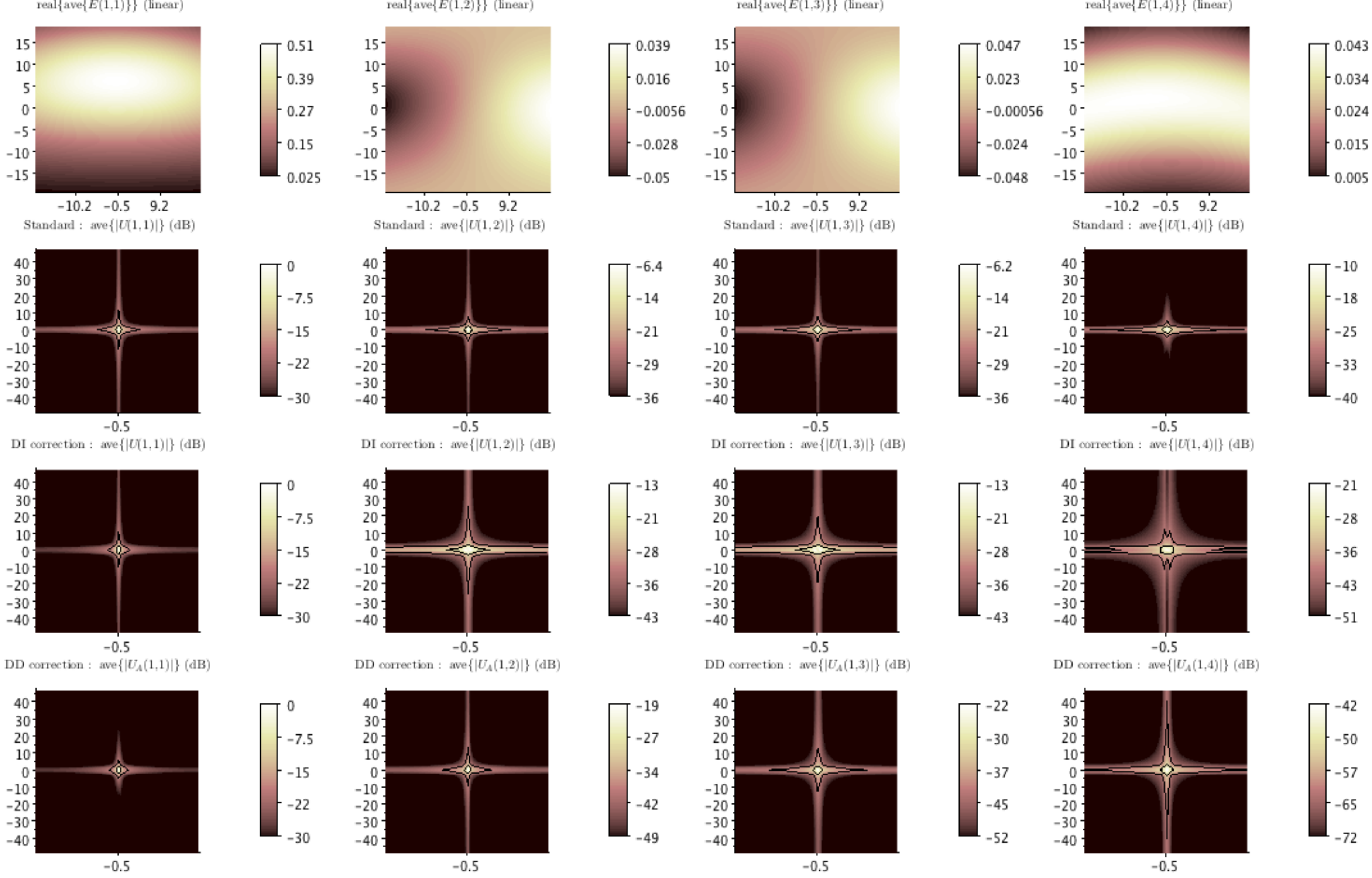}
\end{center}

\caption{The panels show $\sim40^\circ\times40^\circ$ angular response functions and their FFTs for MWA-style antenna
	 tiles, averaged over 7 snapshots for a single baseline (hour angle = [-3,3], declination $\sim-71^\circ$). They
	 demonstrate once again the benefit of moving to an antenna-based coordinate system, as described in section
	 \ref{Estimating Sky Brightness}. Reading down each column, we have elements of $\mat{\E}_k(\s)$,
	 $\mat{\U}_k(\bl)$, $\mat{\U}_k^\prime(\bl)$, and $\mat{\U}_{Ak}(\bl)$. In the image domain, the real components
	 are shown, with linear amplitude scaling and $\s$ in degrees. In the Fourier domain, averages are over the
	 absolute values, functions have logarithmic amplitude scaling (running from -30 dB to 0 dB relative to the
	 panel maximum, all scaled to the maximum of element 1,1), and $\bl$ is in wavelengths. Tile primary beams are
	 modelled as the superposition of 16 dipole angular response matrices with various dipole-based errors, as
	 described in the main text, and these imperfections set the final level of leakage that is achieved in an
	 antenna-based coordinate system, shown in the bottom row of panels. The high level of leakage seen for the
	 sky-based coordinate system, shown in the second row, is due to the structure in $\mat{\C}(\s)$. The second to
	 last row shows averaged kernels for the situation where the 4$\times$4 feed configuration matrix at the centre
	 of each snapshot, $\C(\s_0)$, is applied to the spatial-coherency sample directly, before gridding. This
	 direction-independent (DI) correction allows the removal of $\C(\s_0)$ from the kernels, leading to the
	 reduction that is seen in the off-diagonal amplitudes. This is akin to a parallactic angle rotation for alt-az
	 dishes, however here the correction is only exact at the centre of the image. While not as good as the full
	 direction-dependent (DD) conversion to an antenna-based system, it does reduce the leakage seen in the second
	 row.}

\label{plot_4x4_tile_beams_track_small}
\end{figure}

\begin{figure}[ht]
\begin{center}
\includegraphics[width=16.5cm]{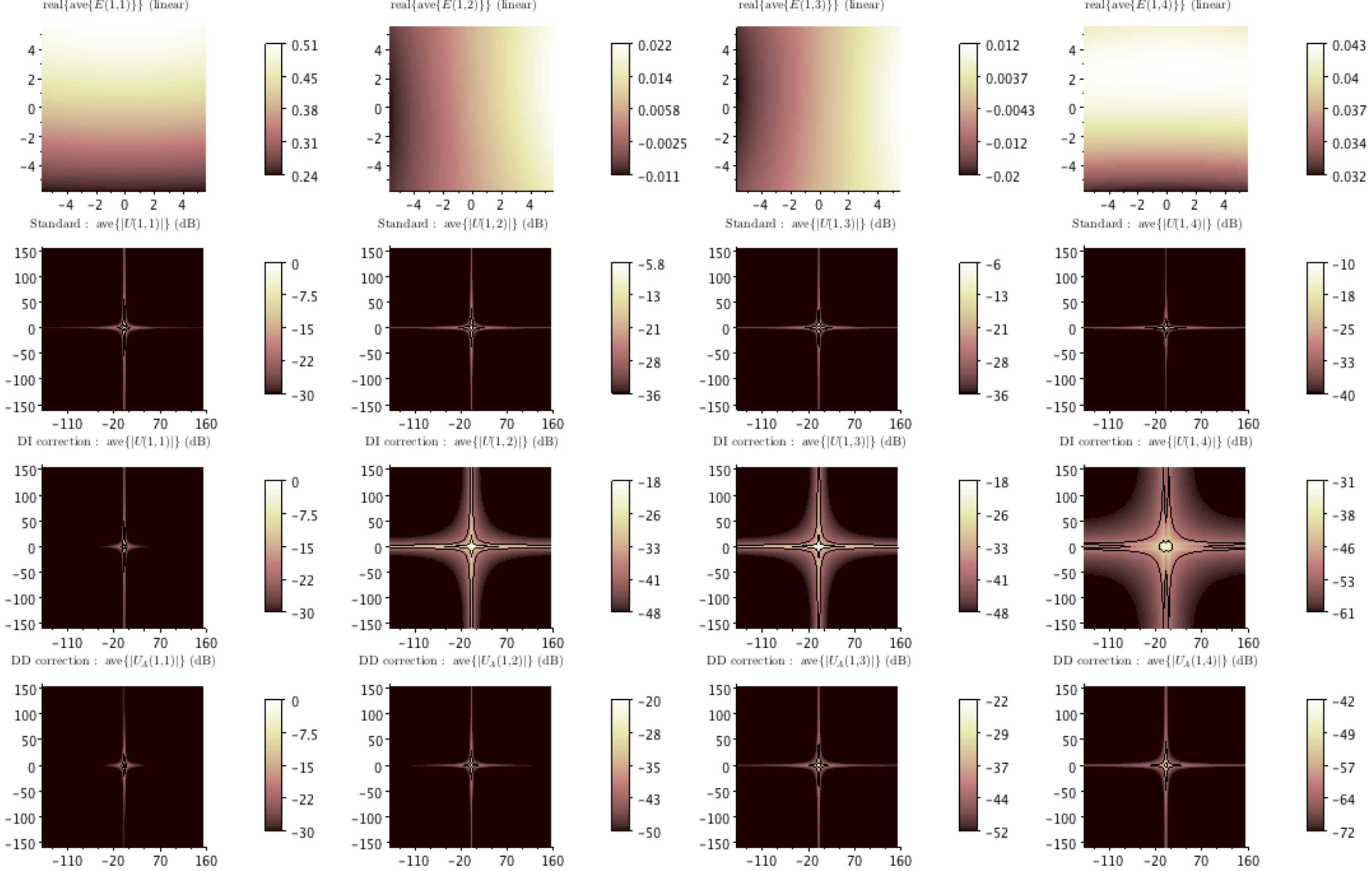}
\end{center}

\caption{The same as figure \ref{plot_4x4_tile_beams_track_small}, however here the image size has been reduced to
	 $\sim12^\circ\times12^\circ$. As in the previous figure, the second to last row shows averaged kernels for the
	 situation where the 4$\times$4 feed configuration matrix at the centre of each snapshot, $\C(\s_0)$, is applied
	 to the spatial-coherency sample directly, before gridding. This direction-independent (DI) correction allows
	 the removal of $\C(\s_0)$ from the kernels. The correction is only exact at the centre of the image, and as
	 expected for a reduced field size, the amplitudes of the off-diagonal terms are closer to those of the full
	 direction-dependent (DD) conversion to an antenna-based system, compared with those of the larger field shown
	 in figure \ref{plot_4x4_tile_beams_track_small}.}

\label{plot_4x4_tile_beams_track_smaller}
\end{figure}

\begin{figure}[ht]
\begin{center}
\includegraphics[width=15cm]{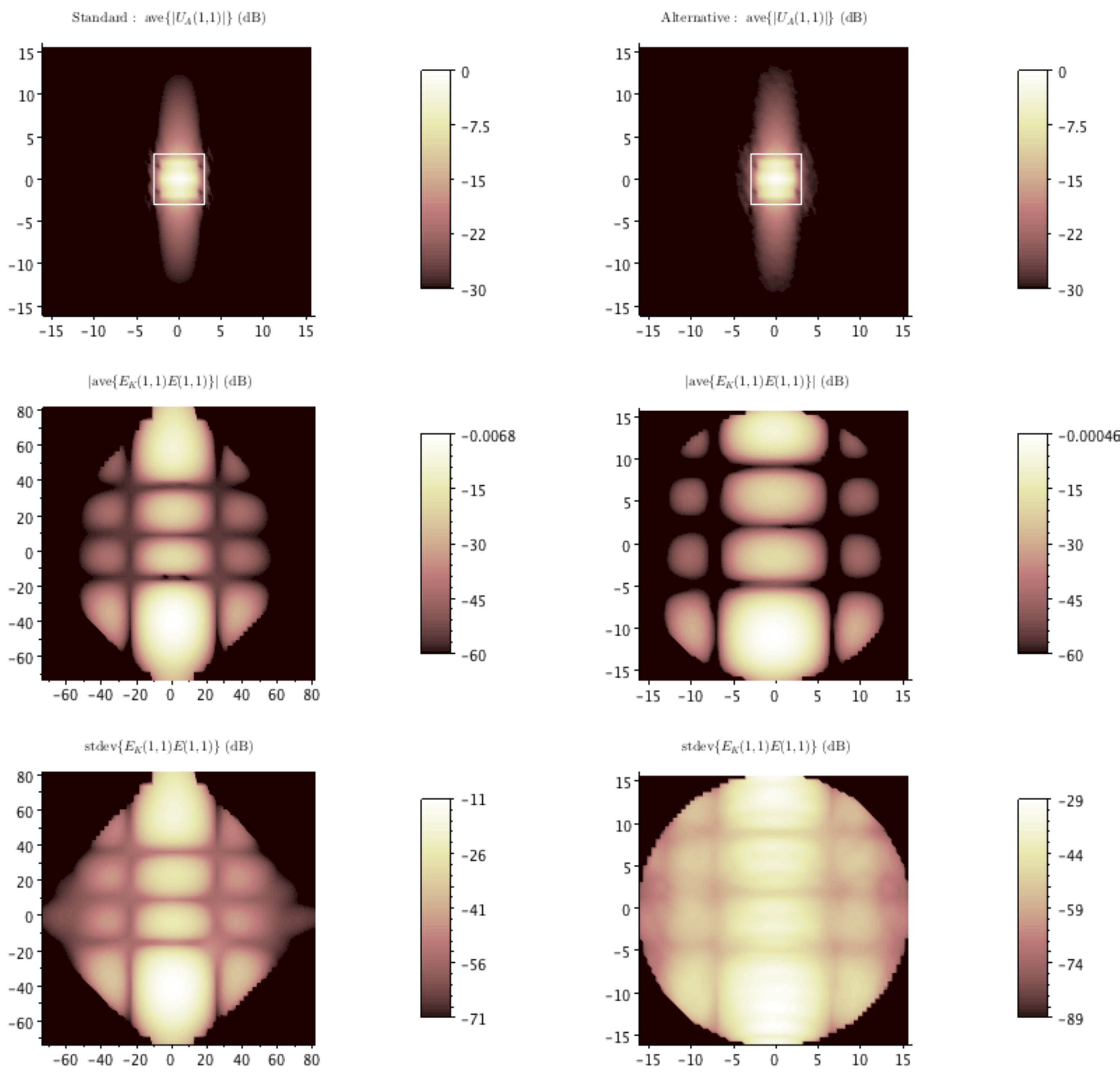}
\end{center}

\caption{The panels show gridding kernels and the associated gridded angular response for the functions shown in figure
	 \ref{plot_4x4_tile_beams_paper}. They demonstrate that the alternative kernels discussed in section
	 \ref{Alternative Gridding Kernels} can have compact support, and the level to which they can suppress
	 differences between the response of the gridded samples. The top left and top right panels show images of
	 $|\Fline{\E_{k,11}^*(\s)}|$ (see section \ref{Estimating Sky Brightness}) and
	 $|\Fline{\bar{\E}_{11}^*(\s)\,\bar{\E}_{11}^{}(\s)/(\E_{k,11}^{}(\s)+\epsilon)}|$ (see section \ref{Alternative
	 Gridding Kernels}), respectively. For the latter, $\epsilon$ was set to 0.1\% of the maximum value of
	 $|\E_{k,11}^{}(\s)|$. The tile primary beams are modelled as the superposition of 16 dipole angular response
	 matrices with various dipole-based errors, as described in the main text and shown in figure
	 \ref{plot_4x4_tile_beams_paper}. The top right panel demonstrates a gridding kernel that divides out the
	 angular response of a sample but retains compact support. Using the regions within the white boxes that are
	 indicated in the top panels to generate gridding kernels, the resulting gridded angular weighting functions,
	 $\E_{K,k}(\s)\E_{k}(\s)$, are shown down each column, where $\E_{K,k}(\s)$ represents the FFT of the cropped
	 region for sample $k$. The middle row shows the average gridded response, the bottom row shows the standard
	 deviation. The sample-to-sample standard deviation after gridding with alternative kernels is almost 20 dB less
	 than the standard deviation after gridding with standard kernels, which will lead to less variation in the PSF
	 and smaller side-lobes (as will uniform or robust weighting).}

\label{plot_4x4_tile_beams_extra}
\end{figure}

Consider a simulated array of crossed dipoles on ground screens, which have the zenith at boresight and dipoles aligned
with North and East in the horizontal coordinate system. Furthermore, suppose the array is at latitude $L$ and that the
dipoles are grouped into MWA-style antenna tiles comprising 16 dipoles. Figures \ref{plot_4x4_tile_beams_paper} through
\ref{plot_4x4_tile_beams_extra} show functions from various elements of the $\mat{\E}_k(\s)$, $\mat{\U}_k(\s)$ and
$\mat{\C}(\s)$ matrices for these antenna tiles. Tile primary beams are modelled as the superposition of the 16 dipole
gain patterns, described below, and they are distorted by the addition of direction-independent Gaussian noise to the
gain and phase of each dipole with an RMS of 10\% and 5$^\circ$ respectively. The dipoles were separated by 1.1 m at a
height of 0.3 m above a ground plane, with Gaussian-distributed position errors of 0.01 m RMS in each direction, and
Gaussian-distributed rotation errors of 1$^\circ$ RMS in azimuth. Gaussian-distributed errors in the boresight of each
tile were also added, with an RMS of 1$^\circ$ from zenith in both directions. While these errors are somewhat
realistic, they should not be thought of as representative for MWA, as they were chosen rather arbitrarily. In direction
$\s$, the feed configuration matrix, $\mat{\C}(\s)$, has been constructed from the tensor product of the 2$\times$2
dipole feed configuration Jones matrix, $\mat{c}(\s)$, with itself, where $\mat{c}(\s)$ converts from the standard
equatorial coordinates of the celestial sphere to the local North and East coordinates:

\begin{equation}
\label{crossed dipole transformation}
\mat{c}(\s) = \left[
\begin{array}{cc}
  \cos{L}\,\cos{\delta(\s)} + \sin{L}\,\sin{\delta(\s)}\,\cos{H(\s)}, & -\sin{L}\,\sin{H(\s)} \\
  \sin{\delta}\,\sin{H(\s)}, & \cos{H(\s)}\\
\end{array}\right],
\end{equation}

where $H(\s)$ and $\delta(\s)$ are the hour angle and declination of direction $\s$ respectively.

Figure \ref{plot_4x4_tile_beams_paper} shows elements of $\mat{\E}_k(\s)$, $\mat{\U}_k(\bl)$, $\mat{\C}(\s)$, and
$\mat{\U}_{Ak}(\bl)$, averaged over $k$. The simulation is for the extreme case of imaging the entire sky with antenna
primary beams that are pointed close to the south celestial pole, as can be seem from the four angular response patterns
in the top row. The second row down shows the effect of describing all of the polarised response information in Fourier
space: the off-diagonal terms have large amplitudes, and much of the information is distributed across Fourier space.
The bottom row shows similar Fourier domain functions, except that the complicated feed configuration functions (shown
in the middle row) were removed, to be dealt with in the image domain. In this antenna-based coordinate system, all of
the elements are seen to have compact support and much reduced off-diagonal amplitudes.

Figures \ref{plot_4x4_tile_beams_track_small} and \ref{plot_4x4_tile_beams_track_smaller} show elements of
$\mat{\E}_k(\s)$, $\mat{\U}_k(\bl)$, and $\mat{\U}_{Ak}(\bl)$, when imaging smaller fields. The panels show the effect
of sidereal motion on the angular response functions of a single sample. In each figure, the second row shows that, once
again, describing all of the polarised response information in Fourier space leads to off-diagonal terms with large
amplitudes. In the third row of each figure, the feed configuration matrix for the centre of the image has been used in
an attempt to align each snapshot in a direction-independent fashion, much like parallactic-angle corrections for alt-az
parabolic dish antennas. The off-diagonal terms are reduced in amplitude\,--\,more so for the smaller field, as one
might expect\,--\,however not as much as after full conversion to the antenna-based system, as shown in the bottom row
of each figure. 

Figure \ref{plot_4x4_tile_beams_extra} gives a comparison of the two gridding approaches discussed in sections
\ref{Estimating Sky Brightness} and \ref{Alternative Gridding Kernels}. It can be seen that, for this type of primary
beam function, both approaches produce kernels with similar support in Fourier space, and the alternative kernel type
discussed in section \ref{Alternative Gridding Kernels} can produce gridded response patterns with reduced variation
from sample to sample.

\section{Summary}

Direction-dependent gridding is reviewed from the perspective of low frequency dipole arrays. It is seen that the
approach discussed by \cite{Bhatnagar-etal.2008} is quite general and can be applied to arrays of dipoles, however
temporal and angular variations in the feed configuration relative to the sky mean that the gridding kernel extent will
likely increase the longer one averages samples in Fourier space.

It is seen that snapshot imaging followed by image-based weighting and averaging can help to keep the amount of
computation involved in visibility gridding to a minimum. Since, at any instant, many current and planned arrays have
baselines that lie close to a two-dimensional plane, the simplifications offered by snapshot imaging couple well with
the warped-snapshot wide-field imaging technique discussed in \cite{Bracewell1984, CornwellPerley1992, SynImRAII-Ch19,
Cornwell-etal.2003}; and \cite{Ord-etal.2010}.

While the maximum-likelihood approach reviewed in section \ref{Estimating Sky Brightness} is in many respects ideal for
making sky images, it is aimed at structure that is above the image noise. Any sky structure that is at the noise
level\,--\,or the noise itself for images the are limited by confusion\,--\,retains the PSF, which in general will be
position, polarisation, time and frequency dependent. This could limit the extraction of science signals that are buried
in the noise, such as the angular power spectrum of redshifted 21cm emission from neutral Hydrogen. We introduced an
alternative gridding scheme in section \ref{Alternative Gridding Kernels}, with the aim of reducing side-lobe structure
by reducing Fourier coefficient structure that is due to the array.


\section*{Acknowledgements}

Support came from the U.S. National Science Foundation (grants AST-0457585 and PHY-0835713). The Centre for All-sky
Astrophysics is an Australian Research Council Centre of Excellence, funded by grant CE11E0090. The International Centre
for Radio Astronomy Research is a Joint Venture between Curtin University and The University of Western Australia,
funded by the State  Government of Western Australia and the Joint Venture partners. R.B.W is supported via the Western
Australian Centre of Excellence in Radio Astronomy Science and Engineering.


\bibliography{CALbib}

\begin{thebibliography}{}

\bibitem[Bernardi et~al., 2011]{Bernardi-etal.2011}
Bernardi, G., Mitchell, D.~A., Ord, S.~M., Greenhill, L.~J., Pindor, B., Wayth,
  R.~B., and Wyithe, J. S.~B. (2011).
\newblock {Subtraction of point sources from interferometric radio images
  through an algebraic forward modeling scheme}.
\newblock {\em MNRAS}, 413(1):411--422.

\bibitem[{Bernardi et al.}, 2009]{Bernardi-etal.2009}
{Bernardi et al.}, G. (2009).
\newblock {Foregrounds for observations of the cosmological 21 cm line. I.
  First Westerbork measurements of Galactic emission at 150 MHz in a low
  latitude field}.
\newblock {\em A\&A}, 500(3):965--979.

\bibitem[{Bernardi et al.}, 2010]{Bernardi-etal.2010}
{Bernardi et al.}, G. (2010).
\newblock {Foregrounds for observations of the cosmological 21 cm line. II.
  Westerbork observations of the fields around 3C 196 and the North Celestial
  Pole}.
\newblock {\em A\&A}, 522:67.

\bibitem[{Bernardi et al.}, 2012]{Bernardi-etal.2012}
{Bernardi et al.}, G. (2012).
\newblock {A 2400 Square Degree Survey at 200 MHz: Bright Sources, Foregrounds
  and Polarization}.
\newblock {In prep.}

\bibitem[Bhatnagar et~al., 2008]{Bhatnagar-etal.2008}
Bhatnagar, S., Cornwell, T.~J., Golap, K., and Uson, J.~M. (2008).
\newblock {Correcting direction-dependent gains in the deconvolution of radio
  interferometric images}.
\newblock {\em A\&A}, 487(1):419--429.

\bibitem[Bowman et~al., 2009]{Bowman-etal.2009}
Bowman, J.~D., Morales, M.~F., and Hewitt, J.~N. (2009).
\newblock {Foreground Contamination in Interferometric Measurements of the
  Redshifted 21 cm Power Spectrum}.
\newblock {\em ApJ}, 695:183.

\bibitem[Bracewell, 1984]{Bracewell1984}
Bracewell, R.~N. (1984).
\newblock {Inversion of nonplanar visibilities}.
\newblock In Roberts, J.~A., editor, {\em {Indirect Imaging}}, volume 177.
  Cambridge: Cambridge University Press.

\bibitem[Cornwell et~al., 2003]{Cornwell-etal.2003}
Cornwell, T.~J., Golap, K., and Bhatnagar, S. (2003).
\newblock {W Projection: A New Algorithm for Non-Coplanar Baselines}.
\newblock {\em EVLA Memo}, 67.

\bibitem[Cornwell and Perley, 1992]{CornwellPerley1992}
Cornwell, T.~J. and Perley, R.~A. (1992).
\newblock {Radio-interferometric imaging of very large fields - The problem of
  non-coplanar arrays}.
\newblock {\em A\&A}, 261:353--364.

\bibitem[de~Vos et~al., 2009]{deVos-etal.2009}
de~Vos, M., Gunst, A.~W., and Nijboer, R. (2009).
\newblock {The LOFAR Telescope: System Architecture and Signal Processing}.
\newblock {\em Proc. IEEE}, 97(9):1431--1437.

\bibitem[Ellingson et~al., 2009]{Ellingson-etal.2009}
Ellingson, S.~W., Clarke, T.~E., Cohen, A., Craig, J., Kassim, N.~E.,
  Pihlstrom, Y., Rickard, L.~J., and Taylor, G.~B. (2009).
\newblock {The Long Wavelength Array}.
\newblock {\em Proc. IEEE}, 97(8):1421--1430.

\bibitem[Furlanetto et~al., 2006]{Furlanetto-etal.2006}
Furlanetto, S.~R., Oh, S.~P., and Briggs, F. (2006).
\newblock {Cosmology at Low Frequencies: The 21 cm Transition and the
  High-Redshift Universe}.
\newblock {\em Physics Reports}, 433:181.

\bibitem[Greenhill and Bernardi, 2012]{GreenhillBernardi2012}
Greenhill, L.~J. and Bernardi, G. (2012).
\newblock {HI Epoch of Reionization Arrays}.
\newblock In Komonjinda, S., Kovalev, Y., and Ruffolo, D., editors, {\em {11th
  Asian-Pacific Regional IAU Meeting 2011}}, volume~1 of {\em NARIT Conference
  Series}. Bangkok, NARIT.

\bibitem[Hamaker et~al., 1996]{Hamaker-etal.1996}
Hamaker, J.~P., Bregman, J.~D., and Sault, R.~J. (1996).
\newblock {Understanding radio polarimetry. I. Mathematical foundations}.
\newblock {\em {A\&A Suppl. Ser.}}, 117:137--147.

\bibitem[{Lonsdale et al.}, 2009]{Lonsdale-etal.2009}
{Lonsdale et al.}, C.~J. (2009).
\newblock The murchison widefield array: Design overview.
\newblock {\em {Proc. IEEE}}, 97(8):1497--1506.

\bibitem[Mitchell et~al., 2008]{Mitchell-etal.2008}
Mitchell, D.~A., Greenhill, L.~J., Wayth, R.~B., Sault, R.~J., Lonsdale, C.~J.,
  Cappallo, R.~J., Morales, M.~F., and Ord, S.~M. (2008).
\newblock Real-time calibration of the murchison widefield array.
\newblock {\em IEEE Journal of Selected Topics in Signal Processing},
  2:1993--2006.

\bibitem[Morales et~al., 2006]{Morales-etal.2006}
Morales, M.~F., Bowman, J.~D., and Hewitt, J.~N. (2006).
\newblock {Improving Foreground Subtraction in Statistical Observations of 21
  cm Emission from the Epoch of Reionization}.
\newblock {\em ApJ}, 648:767.

\bibitem[Morales and Matejek, 2009]{MoralesMatejek2009}
Morales, M.~F. and Matejek, M. (2009).
\newblock {Software holography: interferometric data analysis for the
  challenges of next generation observatories}.
\newblock {\em MNRAS}, 400(4):1814--1820.

\bibitem[Morales and Wyithe, 2010]{MoralesWhiyte2010}
Morales, M.~F. and Wyithe, J. S.~B. (2010).
\newblock {Reionization and Cosmology with 21-cm Fluctuations}.
\newblock {\em Annual Review of Astronomy and Astrophysics}, 48:127--171.

\bibitem[Myers et~al., 2003]{Myers-etal.2003}
Myers, S.~T., Contaldi, C.~R., R.Bond, J., Pen, U.-L., Pogosyan, D., Prunet,
  S., Sievers, J.~L., Mason, B.~S., Pearson, T.~J., Readhead, A. C.~S., and
  Shepherd, M.~C. (2003).
\newblock {A Fast Gridded Method for the Estimation of the Power Spectrum of
  the Cosmic Microwave Background from Interferometer Data with Application to
  the Cosmic Background Imager}.
\newblock {\em ApJ}, 591(2):575--598.

\bibitem[{Ord et al.}, 2010]{Ord-etal.2010}
{Ord et al.}, S. (2010).
\newblock Interferometric imaging with the 32 element murchison wide-field
  array.
\newblock {\em PASP}.

\bibitem[Parsons et~al., 2010]{Parsons-etal.2010}
Parsons, A.~R., Backer, D.~C., Foster, G.~S., Wright, M. C.~H., Bradley, R.~F.,
  Gugliucci, N.~E., Parashare, C.~R., Benoit, E.~E., Aguirre, J.~E., Jacobs,
  D.~C., Carilli, C.~L., Herne, D., Lynch, M.~J., Manley, J.~R., and Werthimer,
  D.~J. (2010).
\newblock {The Precision Array for Probing the Epoch of Reionization: 8 Station
  Results}.
\newblock {\em AJ}, 139(4):1468--1480.

\bibitem[Perley, 1999]{SynImRAII-Ch19}
Perley, R.~A. (1999).
\newblock {Lecture 19. Imaging with Non-Coplanar Arrays}.
\newblock In Taylor, G.~B., Carilli, C.~L., and Perley, R.~A., editors, {\em
  {Synthesis Imaging in Radio Astronomy II}}, volume 180 of {\em Astronomical
  Society of the Pacific Conference Series}, pages 383--400. Astronomical
  Society of the Pacific.

\bibitem[Pindor et~al., 2011]{Pindor-etal.2011}
Pindor, B., Wyithe, J. S.~B., Mitchell, D.~A., Ord, S.~M., Wayth, R.~B., and
  Greenhill, L.~J. (2011).
\newblock {Subtraction of Bright Point Sources from Synthesis Images of the
  Epoch of Reionization}.
\newblock {\em PASA}, 28(1):46--57.

\bibitem[Press et~al., 1986]{Press1986}
Press, W.~H., Flannery, B.~P., Teukolsky, S.~A., and Vetterling, W.~T. (1986).
\newblock {\em {Numerical Recipes: The Art of Scientific Computing}}.
\newblock Cambridge University Press, first edition.

\bibitem[Rau et~al., 2009]{Rau-etal.2009}
Rau, U., Bhatnagar, S., Voronkov, M.~A., and Cornwell, T.~J. (2009).
\newblock {Advances in Calibration and Imaging Techniques in Radio
  Interferometry}.
\newblock {\em Proc. IEEE}, 97(8):1472--1481.

\bibitem[Smirnov, 2011]{Smirnov2011b}
Smirnov, O.~M. (2011).
\newblock Revisiting the radio interferometer measurement equation. ii.
  calibration and direction-dependent effects.
\newblock {\em A\&A}, 527.

\bibitem[{Sullivan et al.}, 2012]{Sullivan-etal.2012}
{Sullivan et al.}, I.~S. (2012).
\newblock {Fast Holographic Deconvolution: A New Technique for Precision Radio
  Interferometry}.
\newblock {In prep.}

\bibitem[{Taylor et al.}, 2012]{Taylor-etal.2012}
{Taylor et al.}, G.~B. (2012).
\newblock {First Light for the First Station of the Long Wavelength Array}.
\newblock {In prep.}

\bibitem[Thompson et~al., 2001]{Thompson-etal.2001}
Thompson, A.~R., Moran, J.~M., and Jr, G. W.~S. (2001).
\newblock {\em {Interferometry and Synthesis in Radio Astronomy, 2nd Edition}}.
\newblock New York: Wiley-Interscience.

\bibitem[{Tingay et al.}, 2012]{Tingay-etal.2012}
{Tingay et al.}, S.~J. (2012).
\newblock {The Murchison Widefield Array: A Square Kilometre Array Precursor at
  low radio frequencies}.
\newblock {In prep.}

\bibitem[{Williams et al.}, 2012]{Williams-etal.2012}
{Williams et al.}, C.~L. (2012).
\newblock {Low Frequency Imaging of Fields at High Galactic Latitude with the
  Murchison Widefield Array 32-Element Prototype}.
\newblock {Submitted to ApJ}.

\end{thebibliography}


\end{document}